\documentclass[sigplan]{acmart}

\usepackage{amsmath,amsfonts}
\usepackage{algorithmic}
\usepackage{graphicx}
\usepackage{textcomp}
\usepackage[]{hyperref}
\usepackage{multirow}
\usepackage{pgfplots}
\usepackage{pgfplotstable}
\usepackage{listings}
\usepackage{caption}
\usepackage{tikz}
\usepackage{tabularx}
\usepackage{enumitem}
\usepackage{subfigure}
\usepackage{subcaption}
\usepackage{float}
\usepackage{booktabs}
\usepackage{siunitx}
\usepackage{pifont}
\usepackage{algorithm}
\usepackage[skins]{tcolorbox}
\usepackage[dvipsnames]{xcolor}
\pgfplotsset{compat=newest}
\usetikzlibrary{pgfplots.groupplots, decorations.pathmorphing}

\definecolor{obsbg}{HTML}{F2F2F2}
\definecolor{green}{HTML}{607456}

\definecolor{sqlcmt}{HTML}{607456}
\definecolor{sqlkw}{HTML}{1F3A5F}

\lstdefinestyle{mfirsql}{
  language=SQL,
  basicstyle=\scriptsize\ttfamily,
  commentstyle=\color{sqlcmt},        
  keywordstyle=\color{sqlkw}\bfseries,
  morekeywords={WITH,QUALIFY,ROW_NUMBER,OVER,PARTITION,MEDIAN,AS},
  morecomment=[l]{--},
  columns=fullflexible, keepspaces=true,
  showstringspaces=false,
  breaklines=false,
  aboveskip=4pt, belowskip=2pt,
  xleftmargin=2pt, framexleftmargin=2pt,
}

\lstdefinestyle{pseudocode}{
  backgroundcolor=\color{white},
  basicstyle=\ttfamily\bfseries\scriptsize,
  breaklines=true,
  captionpos=b,
  commentstyle=\color{ForestGreen},
  keywordstyle=\color{RoyalBlue},
  stringstyle=\color{BrickRed},
  numbers=left,
  numberstyle=\tiny\ttfamily\bfseries\color{black},
  numbersep=7pt,
  frame=single,
  framerule=1pt,
  rulecolor=\color{black},
  showstringspaces=false,
  tabsize=2,
  xleftmargin=2em,
  language=[GNU]C++,
  morekeywords={_mm_clflush, volatile, __attribute__, [[noinline]], MTRACE, MT_PARAM,
                mt_event_context_t, mt_u64, mt_u32, mt_bool},
}

\AtBeginDocument{%
  }

\setcopyright{none} 
\copyrightyear{2026}
\acmYear{2026}
\acmDOI{XXXXXXX.XXXXXXX}

\begin{document}


\newcommand{\Name}{Microflow}

\title{Microflow: Microarchitectural Causal Observability for Deep Cross-Layer Analysis and Optimization}

\author{Saber Ganjisaffar}
\affiliation{%
  \institution{University of California, Riverside}
  \city{Riverside}
  \state{CA}
  \country{USA}
}
\email{sganj003@ucr.edu}

\author{Chengyu Song}
\affiliation{%
  \institution{University of California, Riverside}
  \city{Riverside}
  \state{CA}
  \country{USA}
}
\email{csong@cs.ucr.edu}

\author{Nael Abu-Ghazaleh}
\affiliation{%
  \institution{University of California, Riverside}
  \city{Riverside}
  \state{CA}
  \country{USA}
}
\email{naelag@ucr.edu}

\begin{abstract}

Modern computer architecture relies heavily on simulation to identify bottlenecks and evaluate optimizations. However, existing microarchitectural performance analysis methods are fundamentally limited by an instruction-centric paradigm that captures only downstream symptoms while leaving the true microarchitectural root cause opaque. Because modern processors are governed by complex interactions across non-instruction entities like prefetchers, replacement policies, and shared queue occupancies, instruction-centric frameworks miss the mechanisms that dictate performance. To eliminate this blind spot, we present \emph{Microflow}, a framework that achieves causal observability in microarchitectural simulation. To address this, we introduce the \emph{Microflow intermediate representation (MFIR)}, which models execution through microarchitecture-tailored core abstractions such as flows, resource tenancies, and causal edges. By compiling simulation runs into a relational causal database, Microflow decouples tracing from analytical processing. This transforms complex diagnostics into expressive queries, enabling architects to trace performance symptoms directly to hardware root causes without developing bespoke analysis scripts or running costly re-simulations for every new question. We demonstrate that Microflow solves pathologies opaque to conventional tools. Across CVP-1 benchmarks, Microflow decomposes a 22\% prefetcher oracle headroom by attributing 61.4\% of stall mass to specific hardware prefetcher decisions, yielding a 2.31\% average speedup (peaking at 25.11\%). Furthermore, it exposes the hidden pipeline-blocking residue of wrong-path execution with high portability and precision across simulators.

\end{abstract}



\maketitle

\section{Introduction}
\label{sec:intro}

Computer architecture research has always been quantitatively driven~\cite{hennessy-90}: simulation is used to identify bottlenecks, evaluate new ideas, and validate optimizations. Every design decision, such as a new cache replacement policy, prefetcher, or branch predictor, begins as a hypothesis about \emph{why} the machine underperforms, using simulation to quantify those hypotheses. While simulators have evolved to mirror the complexity of modern processors, the analysis tools that translate raw simulation data into actionable insights have lagged severely behind. Today, this analytical friction is a primary bottleneck in microarchitectural design and optimization.

Existing simulators provide aggregate statistics, often grouping cycles into predefined stall-category taxonomies without even attributing them to the instructions that caused them~\cite{yasin2014top, de2026fyi}. Even tools capable of instruction attribution treat the instruction as the fundamental unit of analysis, relying on sampled per-instruction cycle stacks~\cite{gottschall2023tea, gottschall2024per, desantana2026dip, de2026pesto} or Dynamic Dependency Graph (DDG) critical-path construction~\cite{fields2003interaction, lee2014rpstacks, tanimoto2017dependence, golestani2019performance, golestani2022calipers, bai2023archexplorer, zheng2023ccs, zheng2024dependence}. This narrow focus exposes only the \emph{symptoms} of a performance problem, obscuring the broader system dynamics. As a result, architects are forced to rely on intuition and bespoke analysis scripts to bridge the gap between aggregate statistics and the precise event-level causality needed to explain interactions between arbitrary microarchitectural entities, resources, and mechanisms.

\begin{figure}[t]
  \centering
\includegraphics[width=0.9\linewidth
]{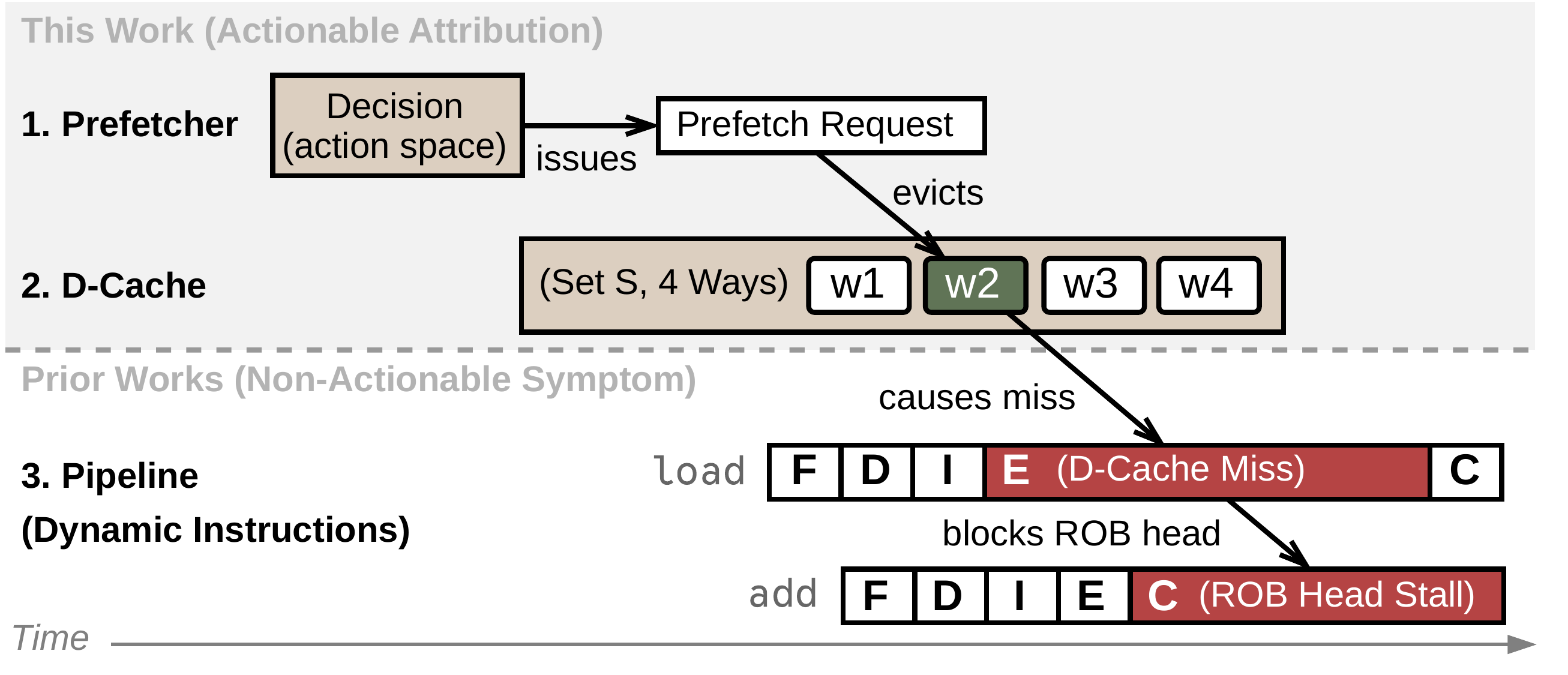}
  \caption{Actionable vs. non-actionable attribution. (Bottom) Prior methods capture local symptoms, such as a stalled load. (Top) Microflow tracks causality to expose the root cause: the prefetch eviction.}
  \label{fig:motiv}
\end{figure}

Consider a scenario where a hardware prefetcher decides to issue a prefetch request, and in doing so, evicts a cache line that a demand \texttt{load} needs a few cycles later. The load misses, reaches the head of the Reorder Buffer (ROB), and stalls an independent, ready-to-commit \texttt{add} instruction behind it. Current analysis methods attribute the delay to the instructions that suffered it: the \texttt{add} stalled due to in-order commit, and the \texttt{load} stalled on a cache miss (Figure~\ref{fig:motiv}, bottom). While technically true, the diagnosis lacks actionable insight because the load is merely the victim rather than the root cause. The actual root cause is the prefetcher's underlying decision to issue the request that evicted the needed line (Figure~\ref{fig:motiv}, top). Because instruction-centric frameworks require an instruction as the anchor for any explanation, they inherently miss causes that are not directly related to instructions (e.g., a prefetch request, a wrong-path load, a page-table walk) and mechanisms that are not direct data dependencies (e.g., eviction, contention, coalescing, queue occupancy). Finding these hidden interactions traditionally requires custom instrumentation followed by costly simulation re-runs. In a complex architecture, these hidden dependencies, which often span multiple structures and non-instruction entities, ultimately determine execution flow and performance. This is not an edge case; as we demonstrate in \S~\ref{sec:bg:motiv}, our analysis reveals that 61\% of L2 demand loads merge into existing prefetch requests, meaning the majority of memory access latencies are dictated by instruction-independent hardware decisions rather than instruction-level dependencies.

These representational blind spots fundamentally limit an architect's ability to extract actionable insights for microarchitectural design and performance optimization. Instruction-centric methods are an effective starting point for locating a bottleneck, but they fall short in explaining its root cause or helping to diagnose a fix. Performance issues do not manifest as abstract instruction stages; rather, they arise from factors such as inaccurate policy \emph{decisions} or contention against \emph{structural limits}. For example, a prefetcher may harm performance because a confidence test issued a polluting candidate; a replacement policy may fail because it selected a victim about to be reused; or a queue may stall the pipeline because its capacity is exhausted by uncommitted speculation. These discrete hardware interactions are entirely hidden by instruction-centric representations. Consequently, a single observed stall behavior can imply opposite hardware interventions, such as tuning a policy versus resizing a buffer. An architect who is told ``37\% of stalls are memory-bound'' has a high-level clue about where to look, but little insight into what is happening or how to mitigate it.

To bridge the gap between an observed performance symptom and the responsible hardware mechanism, every interacting microarchitectural entity, from speculative prefetch requests to shared queue entries, must be explicitly visible to the analysis. Our key insight is that causal observability requires a representational substrate where these non-instruction entities act as explicit connective tissue, directly linking an instruction's delay to underlying hardware decisions, interactions with other actors, and shared resource limits. Preserving these causal interactions enables deeper root-cause analysis and highly targeted architectural optimization. This causal clarity is increasingly critical as the field shifts toward automated, ML-driven hardware design, which requires precise, causative reward signals rather than noisy, aggregate statistics.

Towards this goal, we present \emph{Microflow}, a framework realizing causal observability of microarchitectural simulation. Microflow bridges this gap by operating using an integrated stack for simulation, analysis and optimization of microarchitectural execution. At the event-capture layer, simulators are instrumented using lightweight contracts (\emph{Microtracer}) where causes are recorded exactly where they are decided, rather than inferred offline. By structurally restricting capture to precise regions of interest and specific hardware domains, Microtracer generates focused, high-fidelity causal records without relying on accuracy-degrading statistical sampling. A trace compiler then lifts these raw microarchitectural events into typed entities and identities of the \emph{Microflow Intermediate Representation (MFIR)}. MFIR models the machine through three core abstractions: flows (any unit of work with a lifecycle, such as a dynamic instruction, prefetch request, or cache packet), tenancies (the occupancy of a finite resource by a flow over a time interval), and causal edges. Finally, an analysis engine sits above MFIR, turning the compiled artifact into a relational causal database. Because the microexecution is captured and compiled once, any subsequent diagnostic question, from aggregate bottleneck attribution to precise, cross-entity root-cause traversals, becomes a query executed in seconds. By decoupling simulation from analysis, Microflow eliminates the costly cycle of bespoke re-instrumentation and re-simulation.

We demonstrate the efficacy of MicroFlow through comprehensive case studies that solve complex performance pathologies entirely opaque to existing tools. In \S~\ref{sec:cases:rmax}, we decompose a theoretical 22\% prefetcher oracle headroom into exact causal mechanisms, explicitly attributing 61.4\% of backend-memory stall mass to specific hardware decisions. We prove that suppressed predictions are the true performance bottleneck, even though they produce no nodes in traditional DDG-based methods. This structural insight directly guides a targeted hardware optimization that outperforms prior oracle-driven ablations by over a factor of two. When evaluated on the CVP-1~\cite{perais2018cvp1} benchmark suite, the modification delivers a 2.31\% average speedup across the \textit{compute fp} workloads and reaches a peak improvement of 25.11\%. In \S~\ref{sec:cases:wp}, we trace the hidden structural footprint of speculative wrong-path execution and demonstrate that squashed instructions disproportionately exhaust critical queue capacities exactly when the pipeline blocks.

In summary, this paper makes the following contributions:


\begin{itemize}[leftmargin=1.5em]

\item \textit{\textbf{Achieving Causal Observability}}: We demonstrate that actionable root-cause analysis requires elevating non-instruction hardware entities, internal decisions, and structural limits to first-class analytical primitives.

\item \textit{\textbf{Microflow Intermediate Representation (MFIR)}}: We introduce a novel representation that models microexecution using shared abstractions to preserve complex hardware interactions as explicitly queryable relations.

\item \textit{\textbf{The Microflow Framework}}: We implement a complete compiler stack that decouples simulation capture from analysis. Pairing a capture-time trace with a relational engine turns complex causal traversals into fast software queries.

\item \textit{\textbf{Empirical Demonstration}}: We deploy Microflow to resolve complex performance pathologies, successfully decomposing prefetcher oracle headroom into exact hardware decisions and quantifying the pipeline-blocking structural residue of wrong-path speculation.

\end{itemize}

\section{Background and Motivation}
\label{sec:bg}

\begin{figure}[t]
  \centering
  \includegraphics[width=\linewidth]{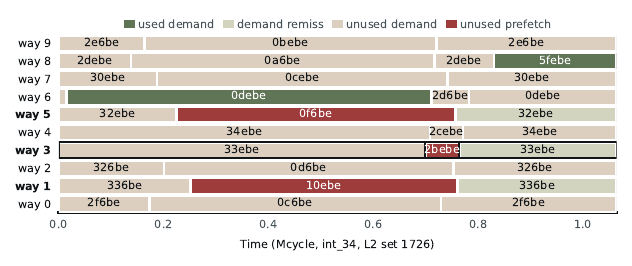}
  \caption{Lifecycle of an L2 cache set. Unused prefetches prematurely evict demand lines, creating a temporal causal disconnect.}
  \label{fig:timeline_motiv}
\end{figure}

To understand why extracting actionable microarchitectural insights requires a representational shift, we must first examine the structural boundaries of existing performance models. Microarchitectural performance analysis has evolved along three dominant lines, each trading off attribution granularity against measurement overhead.

\subsection{Instruction-Centric Attribution}
\label{sec:bg:1}

\textbf{\textit{Aggregate Pipeline-Slot Analysis}}: The most widely deployed methodology relies on low-overhead hardware performance counters to decompose overall execution into coarse bottleneck categories by classifying available pipeline slots. Frameworks based on Top-Down Analysis (TMA)~\cite{yasin2014top} and Interval-based Counter Architectures (ICA)~\cite{eyerman2006performance} evaluate runtime behavior to categorize pipeline activity into high-level performance buckets such as \textit{Frontend Bound}, \textit{Backend Bound}, \textit{Bad Speculation}, and \textit{Retiring}. While commercial implementations are widely adopted across industry~\cite{intelVTune, mundichipparakkal2025arm, amd2024uprof}, recent analyses demonstrate that these architectures lack ground-truth reference validation and incur errors up to $48.9\%$ due to structural misattributions during speculative and back-end stall states~\cite{de2026fyi}. Because these methods reduce microexecution to cumulative slot counters against threshold signals, they yield cheap aggregate breakdowns but fundamentally sacrifice instruction- and event-level causal context.

\textbf{\textit{Per-Instruction Cycle Stacks (PICS)}}: To achieve finer granularity, a second line of work attributes execution time to individual instructions via statistical sampling. Tools like TEA \cite{gottschall2023tea} and DIP \cite{desantana2026dip} establish time-proportional sampling at the commit (egress) and dispatch (ingress) stages, tagging delayed instructions with specific hardware stall reasons (such as a lack of available physical registers or queue entries). Pesto leverages these ingress and egress profiles to discover recurring performance pathologies \cite{de2026pesto}. These techniques offer fine-grained attribution, but they characterize individual instructions in isolation.

\textbf{\textit{Dependence-Graph Critical-Path Analysis}}: The most rigorous approach to root-cause analysis models microexecution as a directed acyclic graph. Nodes represent pipeline events for dynamic instructions, and edges represent data or resource dependencies. By traversing the longest path through this graph, frameworks like Fields et al. \cite{fields2003interaction}, Calipers \cite{golestani2022calipers}, and ArchExplorer \cite{bai2023archexplorer} can pinpoint the exact chain of instructions responsible for total execution time. Recent works have scaled this approach using compressed storage formats \cite{zheng2023ccs} and pattern mining \cite{zheng2024dependence}. This family yields causally grounded bottleneck attribution, but it strictly defines the machine's causal chain as interactions existing only between committed dynamic instructions.

\subsection{Motivation}
\label{sec:bg:motiv}

Modern out-of-order processors rely heavily on hardware-initiated mechanisms to maximize performance, such as prefetchers that generate independent memory traffic and replacement policies that evict lines to manage shared capacity. However, existing performance analysis frameworks share a fundamental limitation: they attribute performance bottlenecks exclusively to the dynamic instruction stream.

To understand how this instruction-centric paradigm breaks down in practice, consider the lifecycle of an L2 cache set from the \texttt{int\_34} benchmark in CVP-1, shown in Figure~\ref{fig:timeline_motiv}. The hardware prefetcher brings in speculative lines (e.g., \texttt{2bebe} in way 3). To accommodate this instruction-independent traffic, the replacement policy actively evicts existing demand lines (e.g., \texttt{33ebe}). Later, the processor demands those exact evicted lines again and suffers costly L2 misses, while the prefetched lines that displaced them are evicted without ever being used (unuseful prefetches). The true root cause of the memory stalls is then the prefetch and eviction event, not the immediate microarchitectural state or the data dependencies of the delayed loads. 

\begin{figure}[t]
  \centering
  \includegraphics[width=\linewidth]{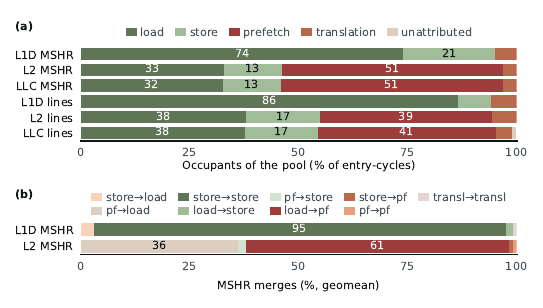}
  \caption{Structural dominance of non-instruction entities. (a) Prefetch requests consume over half of L2/LLC MSHR occupancies. (b) 61\% of L2 demand loads merge into prefetch entries, fundamentally re-routing causality.}
  \label{fig:mshr_motiv}
\end{figure}

This dynamic extends far beyond isolated set conflicts; it governs the behavior of entire structures. Figure~\ref{fig:mshr_motiv} illustrates memory hierarchy occupancies across CVP-1 workloads~\cite{perais2018cvp1} obtained using ChampSim~\cite{gober2022championship}. While the L1 data cache is dominated by demand accesses, the deeper hierarchy reveals a hidden dynamic: at the L2 (where the data prefetcher is enabled) and the LLC, over half of the MSHR occupancy belongs to prefetch requests. When a demand miss encounters an already in-flight request, it merges into the existing MSHR entry. Its completion time is no longer independent; it is dictated by the progress and timeliness of the prefetch occupant. At the L2, a joining demand access merges into a prefetch 61\% of the time. Because these events are generated by the prefetcher rather than individual instructions, instruction-centric tools can only report the final latency, leaving the underlying interaction entirely hidden.

\begin{tcolorbox}[
    enhanced,
    colback=obsbg,
    boxrule=0pt,
    frame hidden,
    borderline west={2pt}{0pt}{green},
    sharp corners,
    left=8pt, right=4pt, top=6pt, bottom=6pt,
    width=\columnwidth
]
\textbf{Observation~1}: Instruction-centric frameworks successfully expose the \emph{symptoms} of performance bottlenecks via delayed instructions, but structurally obscure the underlying \emph{causal mechanisms} required to resolve them.
\end{tcolorbox}

Because prior analytical frameworks structurally bind causality to the instruction stream, this temporal disconnect exposes a severe representational blind spot. Counter-based methods observe aggregate cache miss rates but cannot causally link them to the eviction source. Sampling profilers successfully sample the victim loads but only report their localized memory stalls. Dependence graphs fail entirely: because prefetchers and replacement policies are not dynamic instructions, they cannot exist as nodes in the graph. As a result, the true bottleneck remains structurally invisible to all three families of prior art.

\begin{tcolorbox}[
    enhanced,
    colback=obsbg,
    boxrule=0pt,
    frame hidden,
    borderline west={2pt}{0pt}{green},
    sharp corners,
    left=8pt, right=4pt, top=6pt, bottom=6pt,
    width=\columnwidth
]
\textbf{Observation~2}: Overcoming these structural blind spots requires moving beyond isolated instruction attribution to natively model non-instruction hardware entities and resource occupancies as first-class primitives.
\end{tcolorbox}

This structural failure serves as the primary motivation for MicroFlow. Achieving actionable causal observability requires a new representational substrate that bypasses the instruction-centric boundary, natively modeling non-instruction hardware entities as queryable primitives.
\section{\Name{} Framework Design}
\label{sec:framework}

\begin{figure}[t]
  \centering
  \includegraphics[width=0.94\linewidth]{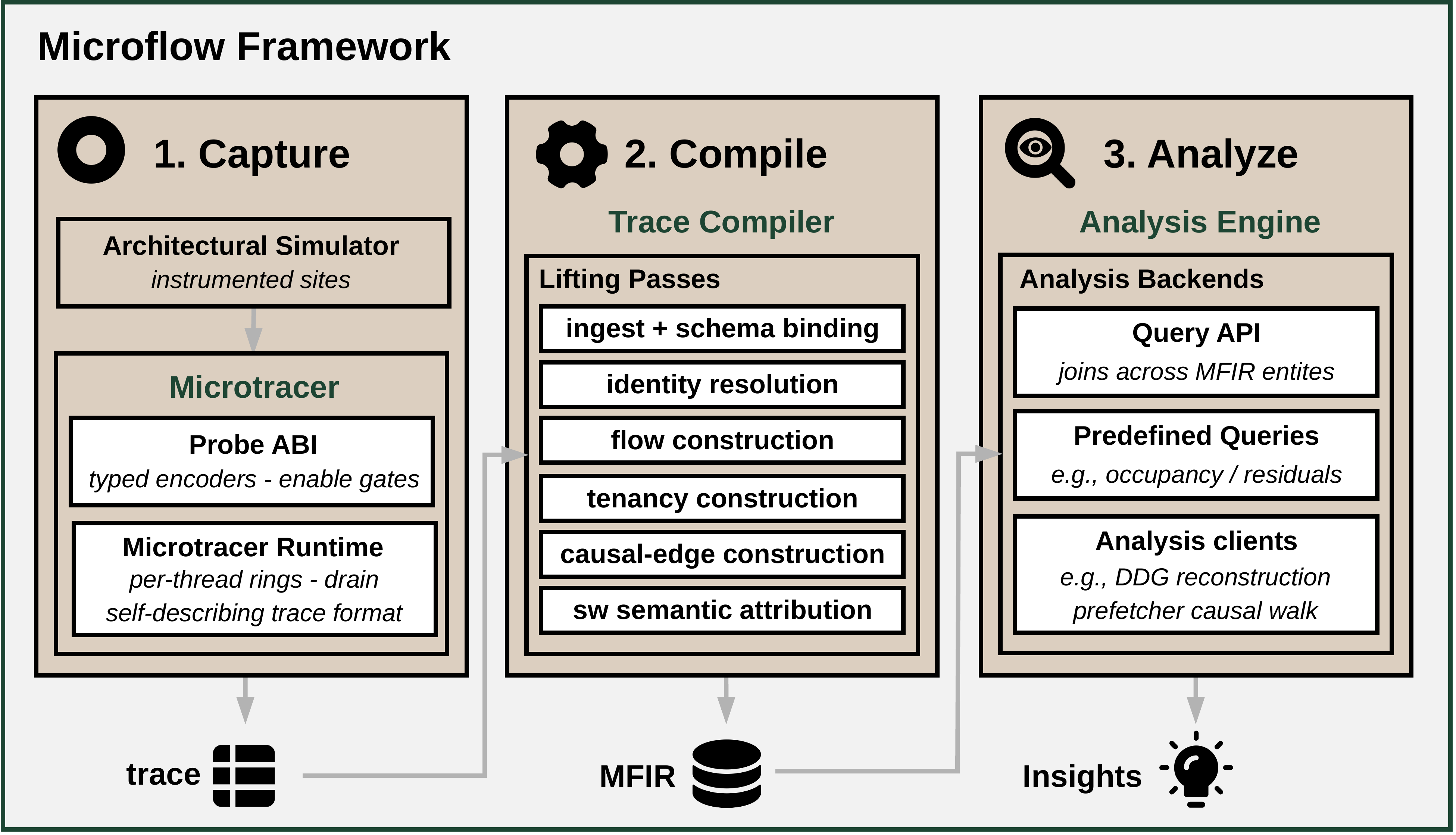}
  \caption{Overview of the Microflow Framework.}
  \label{fig:framework}
\end{figure}
\Name{} operates like a compiler toolchain built around a single intermediate representation of a simulated microexecution (Figure~\ref{fig:framework}), where architects can write reusable and compositional analysis passes against different simulators.
The framework consists of three distinct phases: \emph{1.~Capture}, \emph{2.~Compile}, and \emph{3.~Analyze}. This decoupled pipeline mirrors the design principles of modern software telemetry architectures~\cite{perfetto}. By explicitly separating the fast-path capture from the heavy relational processing, \Name{} ensures that complex identity resolution and causal joins remain entirely out of the simulator's critical path. During simulation, a probe layer called \emph{Microtracer} records microarchitectural events together with the evidence of decisions made at instrumented sites. A \emph{trace compiler} then lifts this raw event stream into the \emph{Microflow Intermediate Representation (MFIR)}. Finally, an \emph{analysis engine} exposes the compiled causal database to various queries and analysis backends, including attribution, causal explanation, critical path, occupancy, and headroom measurements.

The primary design goal is to enable \emph{causal observability} for microarchitectural analysis. Architects can directly query which identity held a resource, which transaction changed shared state, and which constraint linked two events. This eliminates the need to reconstruct relationships using fragile, ad-hoc correlations over raw addresses or timestamps. By compiling a captured trace into MFIR, the framework provides a ``capture once, query many'' paradigm. An architect can execute limitless queries against a single compiled artifact without the high cost of re-instrumenting or re-running the simulator for every new research question.


\subsection{Why an Intermediate Representation}

\begin{figure*}[t]
  \centering
  \includegraphics[width=0.94\textwidth]{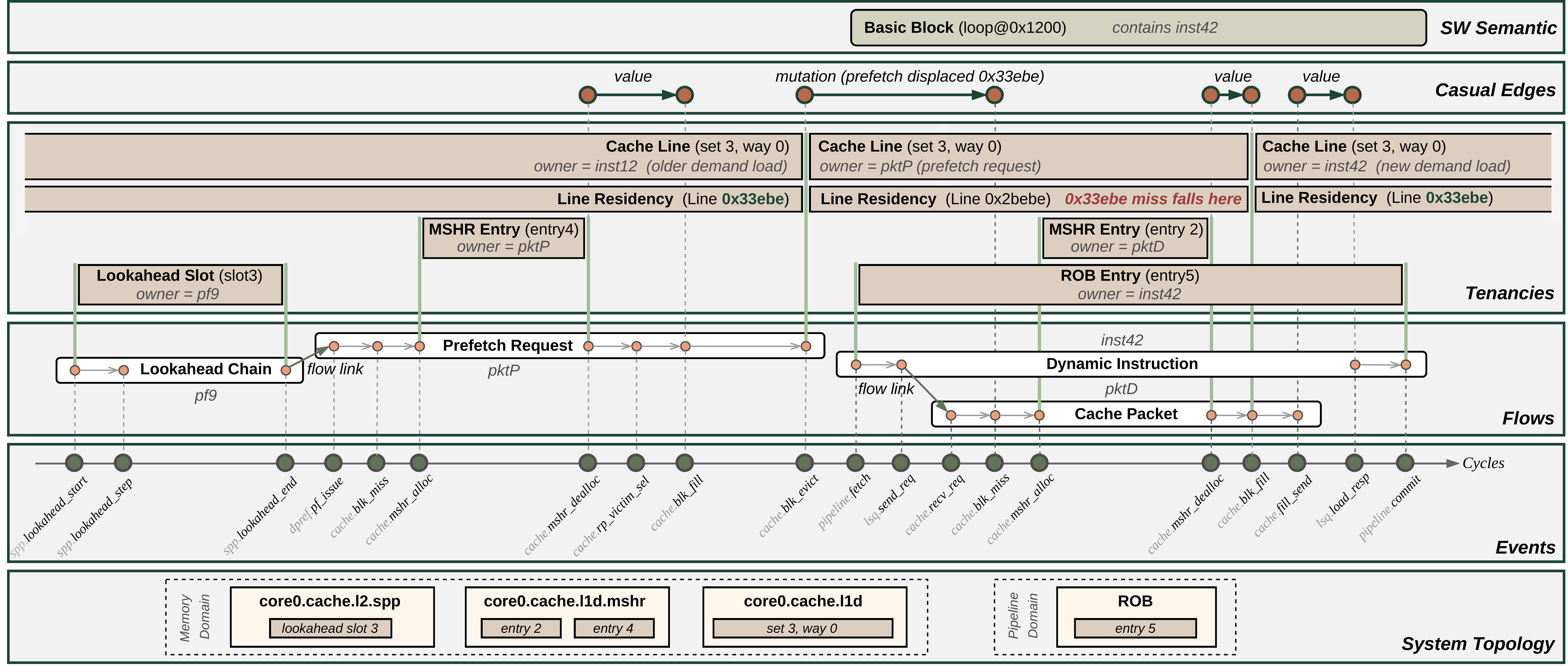}
  \caption{MFIR layered view of prefetch-induced pollution. Events form lifecycled flows; tenancies record resource occupancy; causal edges are measured constraints between events (value dependencies and mutations of shared state). A mutation edge joins the prefetch's displacement of line \texttt{0x33ebe} to the later demand miss.}
  \label{fig:mfir}
\end{figure*}

A raw event trace is notoriously difficult to analyze. Each new research question forces architects to rebuild ad-hoc correlations using addresses, timestamps, or pointer values. These fragile heuristics break under key reuse, multi-level caches, and speculative execution. Conversely, manually crafted dependence graphs are analyzable, but only within the rigid vocabulary their authors initially fixed. 

Furthermore, architectural tooling suffers from severe fragmentation. Architects typically write bespoke analysis scripts tailored to a single simulator to answer a single question. Adapting those scripts to a different simulator or another research question demands a complete rewrite. MFIR solves this fragmentation by adopting the design philosophy of modern compiler infrastructures~\cite{lattner2004llvm, lattner2021mlir}. It serves as a unified middle-end. The trace compiler lifts simulator-local evidence into shared microarchitectural abstractions: \emph{lifecycled entities}, \emph{resource tenancies}, and \emph{typed causal relations}.

By compiling diverse front end simulators into one shared format, MFIR eliminates redundant engineering effort. Supporting a new simulator simply requires mapping its events to these shared schemas. Analysis modules can then query the shared IR directly, completely eliminating the need to rewrite downstream analysis scripts for every new platform.

\subsection{Layered Abstractions and Causality}

MFIR is not defined as a single fixed dependence graph. It is a database of typed tables organized in strict abstraction layers that separate facts that must not be conflated. MFIR adapts proven paradigms from distributed telemetry~\cite{opentelemetry, de2012open} and transaction-level modeling~\cite{systemctlm} to the microarchitectural domain. It reconstructs dynamic hardware flows, such as cache packets and prefetch requests, as first-class entities with strict lifecycles. These flows are no longer treated as mere side effects of instructions. MFIR explicitly models resource occupancy as tenancy intervals against declared capacities. Its core novelty lies in embedding these high-level abstractions into the strict, cycle-accurate, causal reality of microarchitecture. This allows analysis backends to natively join instruction flow, physical resource occupancy, and hardware logic decisions without heuristic guessing.

Figure~\ref{fig:mfir} shows a simplified MFIR for the motivating example from Figure~\ref{fig:motiv}. The database relies on four core entity types. \emph{Events} constitute the immutable ground truth log. \emph{Flows} group these events into lifecycled entities that persist over intervals, with native keys resolved at capture time, such as dynamic instructions, cache packets, and prefetch transactions. \emph{Tenancies} record which flow held a specific resource entry over a specific interval, including owner and coalesced roles when the machine shares an entry. Tenancies reference nodes in the declared \emph{system topology} layer directly, avoiding ad-hoc reconstruction of the capacity or other resource constraints. \emph{Causal edges} represent measured constraints between events, $t(v)\ge t(u)+w$, admitted only when capture evidence supports them. Finally, an optional \emph{software semantic} overlay binds flows to program contexts like basic blocks, functions, or source regions, when that information is available.

Traditional DDG methods typically infer dependency edges offline using a static machine model~\cite{fields2003interaction, lee2014rpstacks, tanimoto2017dependence, golestani2019performance, golestani2022calipers, bai2023archexplorer, zheng2023ccs, zheng2024dependence}. This inference fails quietly when the model and the actual implementation disagree. \Name{} adopts a strict, opposite rule: \emph{Causes are recorded exactly where they are decided and are never reconstructed afterward}. \Name{} captures causality across five channels that recur in synchronous machine models. These include value flow, occupancy, arbitration, speculation boundaries, and state-history mutations of shared structures. Simulators populate these channels precisely where the hardware model stalls a request or updates shared state. The compiler then joins those records on native simulation-time keys. This eliminates the need to infer edges offline. If an analysis requires a new relation, architects simply define a new join over existing identities, or a catalog declaration for a mechanism-specific entity, rather than a new global graph definition.


\subsection{Framework Architecture}

\paragraph{Microtracer.}
Microtracer serves as the capture front end beside the architectural simulator. Event schemas declare typed records for distinct domains like pipelines, caches, or prefetchers. Instrumentation sites call gated encoders to pack fields defined by these schemas, including role-tagged evidence like producer keys, resource keys, acquire and release markers, and structure versions. The runtime writes those records into per-thread buffers and drains them into a self-describing trace file format whose header embeds the schema, ensuring a capture remains readable without the simulator that produced it. Because comprehensive causal tracing requires rich evidence, trace volume inherently scales with execution length. To manage this overhead, MicroTracer employs strict spatial and temporal filtering. Temporally, the runtime dynamically toggles capture based on simulation cycles or instruction counts to target exact Regions of Interest (ROIs). Spatially, architects can selectively trace distinct domains, isolating the memory subsystem from the core pipeline. Furthermore, the runtime can adjust these domain filters on the fly based on simulation signals. For example, it can dynamically disable high-volume cache events after passing a specific instruction milestone. While existing frameworks often rely on sampled tracing to further reduce data volume~\cite{wunderlich2003smarts}, Microflow strictly avoids sampling. Omitting events within an active ROI would break the continuous causal relations required for accurate MFIR analysis. Extending causal reconstruction to tolerate sampled traces is left for future work. Ultimately, the raw stream records evidence rather than finished answers; identity resolution and causality are deferred to the compiler.

\paragraph{Trace Compiler.}
The trace compiler lifts a Microtracer capture into MFIR through an ordered pipeline of passes steered by declarative catalogs rather than by simulator-specific hardcoding. The ingest pass binds the embedded schema and materializes typed event tables. Next, identity resolution replays each event through its declared lifecycle to open, extend, and close flows on native keys instead of heuristics. Subsequent passes build tenancies against the system topology's declared capacities, and join the five causal channels into causal edges whose weights are measured gaps between events. Architecture profiles supply the local nouns and capacities for a CPU target, while the pass structure itself remains shared. The output is a compiled table bundle for one run (events, flows, tenancies, edges, and supporting indexes) that analyses consume directly, rather than
re-deriving identity or occupancy from addresses and timestamps for every question.

\paragraph{Analysis Engine.}
A compiled MFIR artifact serves as a queryable causal database. The analysis engine provides a query layer allowing backends to perform joins across MFIR entities. Clients use SQL API to ask complex identity, occupancy, and causality questions. A classical dependence-graph reconstruction is simply one such client: a \textit{projection} onto instruction-stage vertices under published rules. Another client might generate a cycle-stack report by grouping residuals by instruction flow. \S~\ref{sec:cases:rmax} demonstrates this by building a causal graph for prefetcher performance diagnosis. Mandatory MFIR passes already supply records for dynamic instruction flow, MSHR and cache line residency. The case study simply adds prefetcher-specific entities and decision labels. It then queries the database to join each stall to the exact prefetcher decision that produced it. The same pattern scales to other mechanisms. The facts required for microarchitectural reasoning exist as typed rows, turning explanations into standard short SQL queries rather than bespoke offline correlation scripts as demonstrated in \S~\ref{sec:cases:rmax:attrib}.
\section{Experimental Methodology}
\label{sec:method}

\begin{table}[t]
\centering
\footnotesize
\setlength{\tabcolsep}{4pt}
\begin{tabular}{@{}lll@{}}
\toprule
 & \textbf{gem5 (O3, SE)} & \textbf{ChampSim} \\
\midrule
Clock            & 3.0\,GHz               & 4.0\,GHz \\
Width F/D/X/R    & 8/8/8/8                & 10/6/5/8 \\
ROB/IQ/LQ/SQ     & 192/64/32/32           & 512/205/192/114 \\
Phys.\ regs      & 256 int/fp/vec         & --- \\
Branch pred.     & TAGE-SC-L 64\,KiB      & Hashed perceptron \\
BTB/RAS/FTQ      & 16\,K/16/8             & 8\,K/64/--- \\
L1-I             & 32\,KiB/8/1+1/16       & 32\,KiB/8/4/8 \\
L1-D             & 32\,KiB/8/1+1/16       & 48\,KiB/12/5/16 \\
L2               & 4\,MiB/16/10+10/20     & 1.25\,MiB/10/15/32 \\
LLC              & ---                    & 3\,MiB/24/20/64 \\
Prefetchers      & I: FDP+NL; D, L2: stride & L2: none/SPP/R-Max \\
DRAM             & DDR3-1600, 1\,ch.      & 3200\,MT/s, 1\,ch. \\
Workloads        & 15 SPEC17 (speed)      & 135 CVP-1 traces \\
Warmup/ROI       & 1\,M/2\,M insts        & 20\%/80\% of trace \\
\bottomrule
\end{tabular}
\caption{Simulated configurations for \S\ref{sec:cases:wp} (gem5) and
\S\ref{sec:cases:rmax} (ChampSim, R-Max artifact configuration).}
\label{tab:microarch_config}
\end{table}

All experiments use two cycle-level simulators configured with the parameters in Table~\ref{tab:microarch_config}. For the prefetching case study (\S\ref{sec:cases:rmax}), we use the trace-driven ChampSim simulator~\cite{gober2022championship} as packaged in the R-Max authors' artifact~\cite{wang2026r}, with its configuration unchanged. We run all 135 public CVP-1 traces~\cite{perais2018cvp1} following the authors' protocol: the first 20\% of each trace warms up the caches and predictors, and the remaining 80\% is measured. The no-prefetch baseline, SPP, and R-Max arms differ only in the L2 prefetcher, and the R-Max bound is taken after 12 oracle iterations. For the wrong-path case study (\S\ref{sec:cases:wp}), we use gem5~\cite{lowe2020gem5} with its O3 CPU in System Emulation (SE) mode on 13 SPEC CPU2017 speed benchmarks~\cite{SPEC2017}. SPEC benchmark representative phases are identified with SimPoint methodology~\cite{hamerly2005simpoint}; each benchmark is fast-forwarded to its representative phase checkpoint, warmed up with 10M Atomic instructions, then simulated in detailed O3 CPU mode for 10M committed instructions.
\section{Evaluation}
\label{sec:eval}

In this section, we evaluate \Name{}'s analytical power and foundational properties across three areas. We first apply \Name{} to two complex domains: causal decision tracking for hardware prefetchers (\S\ref{sec:cases:rmax}) and cross-domain structural joins for speculative execution residue (\S\ref{sec:cases:wp}). Finally, we evaluate the framework's scalability and portability across independent simulators (\S\ref{sec:eval:xsim} and \S\ref{sec:eval:overhead}).

\subsection{Case Study~1: Explaining an Oracle's Headroom}
\label{sec:cases:rmax}

R-Max~\cite{wang2026r} is a framework that approximates the realistic upper bound of cache prefetching and replacement mechanisms. It combines perfect knowledge of future memory accesses with strict physical constraints, such as bandwidth, capacity, and MSHRs. The framework iteratively applies an extended B\'el\'ady's MIN algorithm to memory traces to determine optimal prefetch timing without violating these hardware limits. Suppose an architect runs R-Max against the prefetcher they ship and learns that an oracle would deliver a geometric mean IPC improvement of 22\%. Such a bound becomes actionable only if it can be translated into a specific design decision. The architect must establish which fraction of the reported headroom is reachable, and which mechanism must change to reach it. Neither question is answered by the bound itself. 

We demonstrate that a causal record answers both questions. This case study evaluates the Signature Path Prefetcher (SPP)~\cite{kim2016path}, which is one of the prefetchers evaluated in R-Max paper. The study is scoped to the prefetching half of the bound, because prior work reports that the replacement half contributes little to the CVP-1 headroom~\cite{wang2026r}. \textbf{Our goal is to identify the specific decisions within SPP that produce the reported gap, and then to establish that one such attribution is correct.}

\subsubsection{Quantifying the Oracle Headroom}

\begin{figure}[t]
  \centering
\includegraphics[width=0.9\linewidth
]{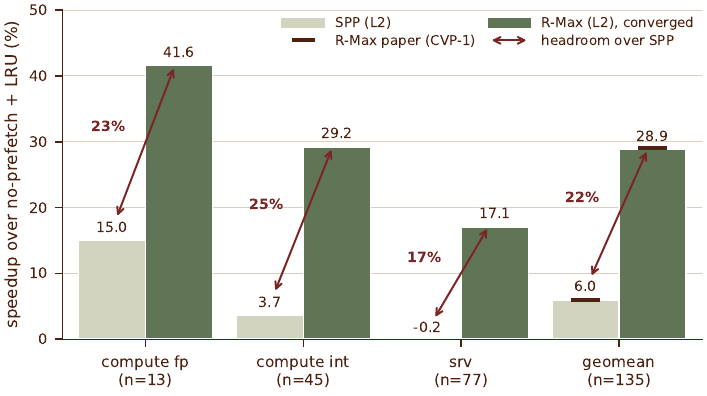}
  \caption{SPP and R-Max on CVP-1, both as speedup over a no-prefetch LRU baseline.}
  \label{fig:rmax_reprod}
\end{figure}

We reproduce the R-Max bound using the authors' artifact~\cite{wang2026r} and 135 CVP-1 traces~\cite{perais2018cvp1}. All speedups are measured against a baseline using no prefetcher and LRU replacement. The SPP arm differs from that baseline solely by adding SPP at the L2.  Figure~\ref{fig:rmax_reprod} presents the converged headroom. Relative to SPP, R-Max yields 23\% on \textit{compute fp}, 25\% on \textit{compute int}, and 17\% on \textit{srv}, resulting in an overall gap of 22\%.

\paragraph{Anchoring the Gap in the Record.}
Attributing this bound requires an additive quantity so no cycle is double-counted. MFIR supplies one: recorded episodes where the ROB head is blocked on memory. Because there is only one ROB head, these episodes are disjoint by construction. Head-blocked memory time constitutes 68.1\% of average execution. Each episode is then attached to the stalling line via persistent flow identities. By linking the load's wakeup directly to the miss that produced it, the query bypasses fragile temporal address-matching. Every percentage in the decision breakdown below is a share of this episode mass; percentages of the R-Max headroom itself are labeled as such.

\subsubsection{Attributing Stalls by Decision}
\label{sec:cases:rmax:attrib}

\begin{figure}[t]
  \centering
\includegraphics[width=\linewidth
]{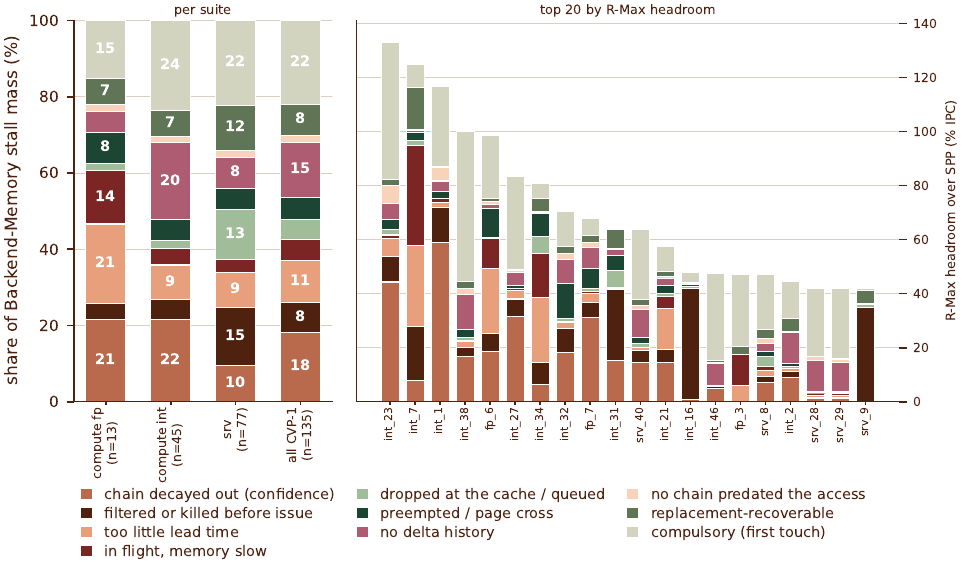}
  \caption{Stall cycles attributed to precise SPP lookahead failure mechanisms.}
  \label{fig:rmax_decisions}
\end{figure}

MFIR can attribute stalls at any granularity. The most intuitive grouping is by the prefetch outcome: never prefetched, dropped before issue, issued too late, or filled but evicted early. Grouping stalls purely by outcome reveals that 82\% of the gap stems from addresses that were never prefetched or were dropped before issue. However, outcome names a state, and a state admits several mechanisms. This broad categorization implies a straightforward solution: \textit{increase prefetch coverage}. Lowering SPP's fill threshold from 25 to 6 yields a negligible +0.14\% overall IPC gain because a single global threshold cannot simultaneously fix diverging root causes. Attribution by \emph{decision} recovers this distinction.

SPP predicts a load's next address from a sequence of previously observed address deltas. Each access advances a per-page lookahead chain. The current signature indexes a pattern table entry, which returns a candidate delta and a confidence value. If this confidence clears the fill threshold, the candidate issues as a prefetch and the chain steps one line deeper. Otherwise, the chain terminates. Termination also occurs if a chain crosses a page boundary or encounters an untrained signature. Figure~\ref{fig:rmax_decisions} categorizes stalls based on exactly where this sequence failed. The breakdown is produced by a single, automated query cascade applied unchanged across all workloads, asking three exact questions of each stalling line:

\begin{enumerate}
    \item \emph{Did a prefetch request exist for this line?} If a request was issued, the query evaluates its runtime lifecycle: whether it merged into an outstanding request (\textit{too little lead time}), was dropped due to structural queue limits (\textit{dropped at the cache / queued}), remained in flight under slow memory (\textit{in flight, memory slow}), or was filled and subsequently evicted before the demand arrived (\textit{replacement-recoverable}).
    \item \emph{If no request was issued, was the address evaluated and suppressed (Active Evaluation)?} If the prefetcher actively calculated the target address but withheld it, recorded decision fields isolate the exact policy barrier: filter rejections, queue capacity limits, confidence failures (\textit{filtered or killed before issue}), or page boundary rules.
    \item \emph{If no evaluation occurred, how far did the lookahead chain reach (Traversal Limits)?}  Where a chain was walking the page, its termination record names why it stopped short: confidence decay (\textit{chain decayed out}), preemption by a new trigger, a page boundary, or an exhausted pattern table. Where no usable chain existed, the episode falls instead to the untrained tail (\textit{no delta history}) or, when nothing in the window could have predicted the line, to \textit{compulsory}.
\end{enumerate}

Listing~\ref{lst:chainstop} answers the third question by addressing the scenario carrying the largest share of stall mass, in which a lookahead chain was already active on the page when the line missed, allowing the record to reveal its specific termination reason. This case accounts for 24\% of Backend-Memory stall mass, and the query resolves it entirely in a single pass applied uniformly across all workloads. The query evaluates only recorded fields (the page traversed by the chain, its start cycle, and its termination reason), and the only non-trivial predicate excludes chains triggered by the missing access itself, since those began after the line was already absent and therefore cannot explain its absence.

\begin{figure}[t]
\begin{lstlisting}[style=mfirsql]
-- Charge each ROB-head memory stall to the SPP chain
-- that gave up before reaching the missing line.
WITH nearest AS (
  SELECT s.service, c.uid
  FROM   head_stall      s   -- disjoint episodes
  JOIN   lookahead_start c   -- chains on that page
    ON   c.page  = (s.line >> 6)   -- 64B line, 4KB
   AND   c.cycle <  s.start_cycle  -- chain came first
   AND   c.started                 -- else uid is 0
   -- exclude the chain this very miss triggered:
   -- it began after the line was already absent
   AND  (c.address >> 6) <> s.line
  -- newest surviving chain; uid breaks cycle ties
  QUALIFY ROW_NUMBER() OVER (PARTITION BY s.stall_row
          ORDER BY c.cycle DESC, c.uid DESC) = 1)
SELECT e.final_outcome       AS why_chain_stopped,
       MEDIAN(e.final_depth) AS steps_taken,
       SUM(n.service)        AS stall_cycles
FROM   nearest n
JOIN   lookahead_end e ON e.chain_uid = n.uid
GROUP BY 1 ORDER BY stall_cycles DESC;
\end{lstlisting}
\caption{The third question, as one query; isolating unpredicted misses via \texttt{head\_stall} and mapping them to preceding chain terminations. Output appears in Table~\ref{tab:chainstop}.}
\label{lst:chainstop}
\end{figure}

\begin{table}[t]\centering\footnotesize
\begin{tabular}{lrrr}
\toprule
why\_chain\_stopped & steps & \% chains & \% stalls \\
\midrule
LOW\_CONFIDENCE & 1 &  75.9 & 18.2 \\
PREEMPTED       & 1 &  13.7 &  3.3 \\
PAGE\_CROSS     & 0 &   9.3 &  2.2 \\
PATTERN\_MISS   & 0 &   1.2 &  0.3 \\
\midrule
total           &   & 100.0 & 24.0 \\
\bottomrule
\end{tabular}
\vspace{1em}
\caption{Output of Listing~\ref{lst:chainstop} over 135 CVP-1 captures. \emph{\%~chains} is the share among episodes whose chain stopped short; \emph{\%~stalls} the share of Backend-Memory stall mass, the quantity Figure~\ref{fig:rmax_decisions} plots.}
\label{tab:chainstop}
\end{table}

Table~\ref{tab:chainstop} shows why a global threshold adjustment fails. Its four rows correspond to the two families illustrated in Figure~\ref{fig:rmax_decisions} for active page traversal: \textit{chain decayed out} at 18\% and \textit{preempted / page cross} at 6\%, which together total the 24\% shown in the table's total row. The remaining stall mass lacks a prior chain to inspect. This includes untrained pages (\textit{no delta history} at 15\%), compulsory first touches (22\%), and a 2\% residue where the only chain was triggered by the missing access itself. Confidence decay alone accounts for 18\% of stall mass, making up three-quarters of that 24\% group, and appears on all but two of the 135 workloads. A companion query locates this mass by showing that chains die at a median confidence of 17 against a fill threshold of 25, and on no workload does that median reach 24. Only the top decile of deaths ($p_{90}=23$) lies within two points of the threshold, so a marginal threshold reduction can rescue at most ten percent of this mass, matching the +0.14\% geomean gain found in the sweep. Meanwhile, the \textit{steps} column shows that these chains advanced a median of only one step before dying. The bottleneck is therefore per-step confidence decay rather than insufficient chain length.

Each leaf carries a \emph{causal verdict}: whether a prediction was explicitly killed, whether a chain never reached the line, or whether nothing could have predicted it. The resulting breakdown identifies that insufficient lead time limits \textit{compute fp} (21.0\%), absent delta history limits \textit{compute int} (20.2\%), and predictions \textit{filtered or killed before issue} limit \textit{srv} at 15\%. Furthermore, \textit{chain decayed out} reaches 21\% to 22\% on the compute suites but drops to 10\% on \textit{srv}, where mass shifts toward suppression and cache admission (13\%). A pooled average obscures all three targets. Overall, the query cascade attributes 61.4\% of backend-memory stall mass to a specific prefetcher or cache decision. The remainder are not decisions at all; compulsory first touches, untrained pages (36.6\%), and unresolved cases (2.0\%).

Those two readings select a remedy. If a chain cannot be made to survive longer, a single active lookahead slot yields roughly one prefetch per trigger and idles between them, so prefetch throughput is recovered by running several short chains concurrently rather than by lengthening one; concurrency also attacks the largest family on \textit{compute fp}, \textit{too little lead time}, because more live chains issue their predictions earlier. We therefore replaced SPP's single active lookahead slot with four concurrent chains. On \textit{compute fp}, the suite where those two families carry the most mass, the modification yields a 2.31\% relative speedup, peaking at 25.11\% on the suite's best workload.

The R-Max authors provide an independent ablation, called SPP-Max, that corroborates this diagnosis~\cite{wang2026r}. SPP-Max replays only the addresses that SPP actually issued, so it can evaluate the timing and survival of approved prefetches but not predictions that were never made. SPP-Max improves upon SPP by only 0.36 percentage points. In contrast, our multiplicity modification gains 0.83 percentage points. It achieves this by targeting the exact mechanisms SPP-Max ignores: \textit{chains that decayed or were preempted}. This demonstrates that the performance gap stems primarily from suppressed predictions, rather than the timing of issued ones. While aggregate speedups confirm that our attribution identifies profitable design targets, IPC gains mix multiple interacting mechanisms. Proving true causality requires isolating and reversing an individual decision to observe its direct effect on a single stall, which we examine next.

\subsubsection{Verifying a Single Attribution}

\begin{figure}[t]
  \centering
\includegraphics[width=0.84\linewidth
]{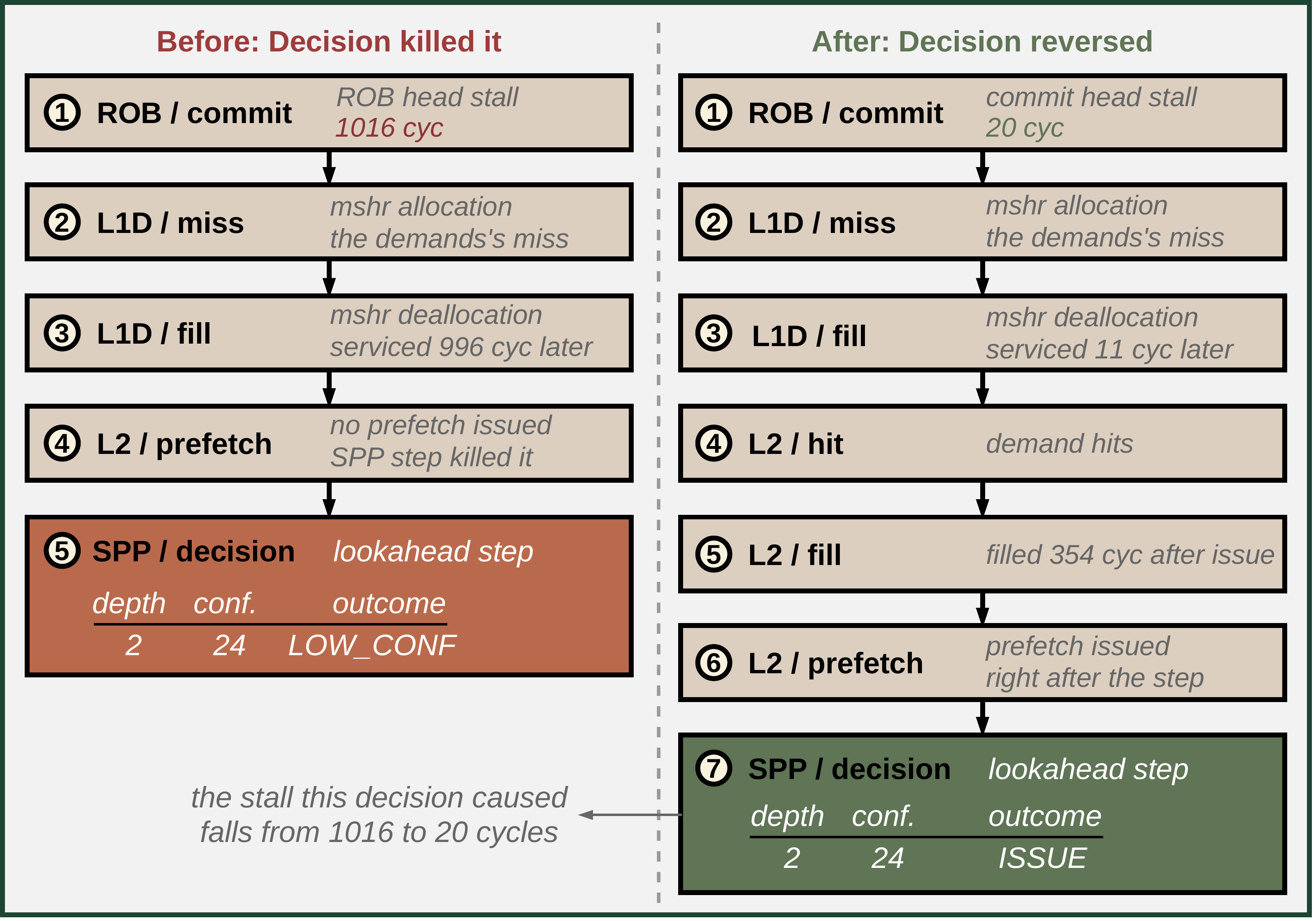}
  \caption{Causal walk of a single SPP prefetch decision before and after targeted reversal. MFIR correctly traces a 1016-cycle ROB stall directly to a suppressed lookahead step, isolating the exact hardware mechanism.}
  \label{fig:rmax_walk}
\end{figure}

To verify MFIR's accuracy in capturing the exact hardware logic behind an omission, we query a single confidence-kill episode backward from a ROB-head stall, through the L1 miss, directly to the exact SPP lookahead step that fell one point short of the confidence threshold (Figure~\ref{fig:rmax_walk}, left). If the attribution is correct, reversing this isolated decision must eliminate the miss. A global threshold change cannot test this, as it alters the running accuracy state. Instead, we reverse only the decision itself, admitting the single suppressed address while changing nothing else, and re-simulate the identical window. As Figure~\ref{fig:rmax_walk} (right) demonstrates, the step now issues, the prefetch arrives early, and the 1016-cycle stall collapses to 20 cycles. This targeted reversal exposes the exact interaction cost~\cite{fields2003interaction}: recovered cycles are offset by longer stalls elsewhere due to displacement, leaving net IPC unchanged. An aggregate metric cannot separate these effects, and a dependence graph misses them entirely because the suppressed decision produced no dynamic instruction node. Accurate causal attribution and an optimized aggregate design change are therefore separate claims; MFIR isolates the former so an architect can engineer the latter.

\paragraph{Scope of the Guarantee.}
Across the 135 workloads, MFIR lands 61.4\% of Backend-Memory stall mass on a named prefetcher or cache decision. The remainder is not a decision at all: 36.6\% is compulsory first touches together with pages SPP had never trained on, and 2.0\% resolves to nothing. Within that 61.4\% the strength of the claim varies, and the record says which is which. Point reversal verifies 7.9\%: predictions the record shows were derived and then killed, where admitting the single suppressed address is a counterfactual the capture licenses. A further 23.3\% is traceable by flow rather than by reversal, because a request existed and its fate is followed from issue to fill or eviction, and 4.5\% is charged to the replacement decision that displaced the line. The last 25.7\% is bounded rather than reversed: a chain stopped before reaching the line, so the hardware computed nothing to reverse, and MFIR reports these as reachability limits rather than algorithmic rejections.

\subsection{Case Study~2: Attributing Wrong-Path Residue}
\label{sec:cases:wp}

A branch misprediction does not end when Wrong-Path (WP) instructions are squashed. Speculative traffic can still occupy shared microarchitectural state that correct-path (CP) execution pays for, or benefits from, after recovery: a Reorder Buffer (ROB) or Issue Queue (IQ) slot held until recovery, a load/store-queue (LQ/SQ) entry, or a demand miss-status
holding register (MSHR) opened by a squashed instruction. Top-Down analysis and IPC show that speculation is expensive; they cannot name \textit{which} shared resource was held, \textit{which} dynamic instruction or prefetch packet held it, or \textit{which} misprediction episode that occupancy belonged to. The goal of this case study is to show that MFIR's typed record of resource occupancy and squash epochs makes that attribution possible across SPEC CPU2017 benchmarks on an out-of-order CPU model simulated in gem5.

\begin{figure}[t]
  \centering
  \includegraphics[width=\linewidth]{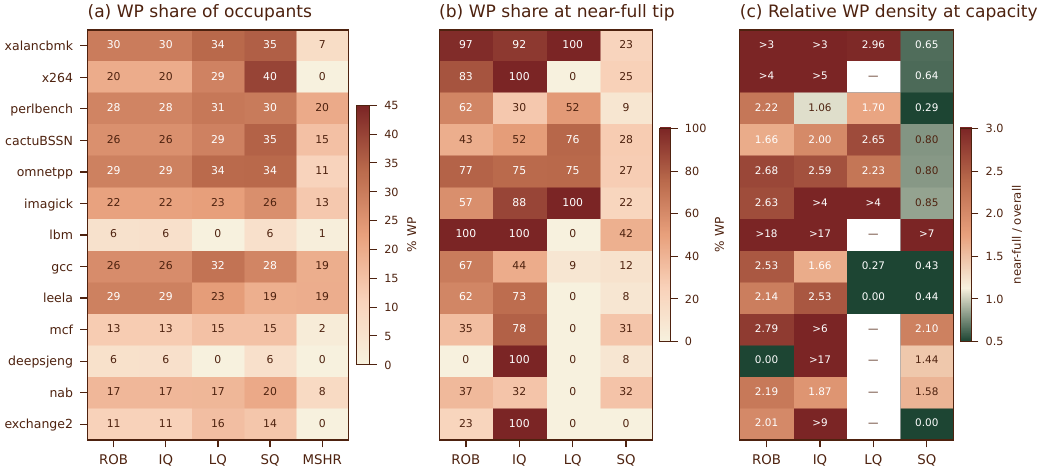}
  \caption{(a)~WP share of occupants.
    (b)~WP share at the near-full tip.
    (c)~Relative WP density (tip$/$overall).}
  \label{fig:wp_residue_occupancy}
\end{figure}

\subsubsection{Occupancy Holders and Squash Epochs}
MFIR records each structure insert and each cache fill as an occupancy of a named entity (a dynamic instruction, or a prefetch packet with no instruction identity). A squash epoch groups the discarded instruction identities of one recovery. Joining occupancy holders to that epoch is what labels a slot or fill as WP residue; counters that only tally full events or misses cannot perform that join. Figure~\ref{fig:wp_residue_occupancy} presents the join as an occupancy footprint. Figure~\ref{fig:wp_residue_occupancy} (a) reports the overall WP share of ROB, IQ, LQ, and SQ inserts (exact instruction identity) and of demand L1D MSHR allocations (packet joined to the memory send, then to the squash epoch).
Figure~\ref{fig:wp_residue_occupancy}~(b)--(c) ask a sharper question: is WP over-represented \textit{when the structure is about to block}? We call those inserts the \emph{near-full tip}: an insert whose recorded occupancy is already at or above 90\% of the observed capacity for ROB/IQ, or within one entry of capacity for LQ/SQ. Figure~\ref{fig:wp_residue_occupancy}~(b) is the WP share among only those tip inserts; Figure~\ref{fig:wp_residue_occupancy}~(c) is the \textit{relative WP density at capacity}, the tip WP fraction divided by the overall WP fraction from Figure~\ref{fig:wp_residue_occupancy}~(a). A value above~1 means WP is denser among capacity-critical inserts than among inserts in general (the color scale is capped at~$3\times$; larger ratios are annotated).

\subsubsection{From Occupancy to Cost}

\begin{figure}[t]
  \centering
  \includegraphics[width=\linewidth]{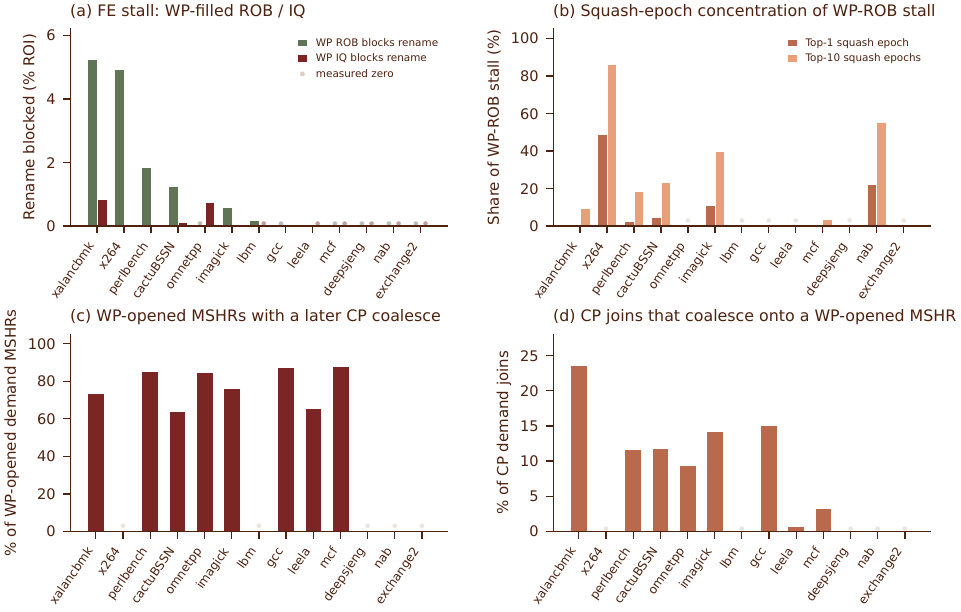}
  \caption{(a)~WP ROB/IQ rename stall (\% ROI).
    (b)~Squash-epoch concentration of WP-ROB stall.
    (c)~WP-opened MSHRs with a later CP coalesce.
    (d)~CP joins onto WP-opened MSHRs.}
  \label{fig:wp_residue_cost}
\end{figure}

Figure~\ref{fig:wp_residue_cost} converts the ROB/IQ cells into a measurable CP cost and situates demand-MSHR residue beside it. Figure~\ref{fig:wp_residue_cost}~(a) charges rename/dispatch as blocked for every cycle where ROB or IQ is at capacity and removing the resident WP holders would free a slot, the frontend back-pressure path from WP occupancy. Figure~\ref{fig:wp_residue_cost}~(b) attributes that WP-ROB stall mass to the squash epochs of the resident WP holders and reports how much of it sits in the top-1 and top-10 epochs. Figure~\ref{fig:wp_residue_cost}~(c)--(d) use the same WP/CP labeling on demand MSHRs, but read a different consequence: a WP instruction may be squashed while its miss remains in flight. Figure~\ref{fig:wp_residue_cost}~(c) asks what fraction of WP-opened demand MSHRs later receive at least one \emph{CP coalesce} (a CP demand packet joining that MSHR). Figure~\ref{fig:wp_residue_cost}~(d) asks the dual share: of all CP demand joins, what fraction land on an MSHR that a WP request opened. Together, Figure~\ref{fig:wp_residue_cost}(c)--(d) quantify in-flight miss reuse from WP to CP, not a frontend stall, but a join that counters cannot attribute to a discarded opener.

\paragraph{Evidence.} 

Figure~\ref{fig:wp_residue_occupancy}(a) shows that WP residue is spread across resources rather than concentrated in one structure. On mispredict-heavy workloads such as \texttt{xalancbmk}, \texttt{omnetpp}, and \texttt{perlbench}, ROB, IQ, LQ, and SQ often carry about 20--35\% WP occupants, while demand MSHRs are usually cooler, with about 7--20\% WP among demand allocations. Conditioning on the near-full tip changes that assessment. Panel~(b) shows that when ROB or IQ is already near capacity, arriving inserts are often majority-WP (for example, greater than 80\% on \texttt{xalancbmk} and \texttt{x264}, and approaching 100\% on several ROB/IQ cells), even though the overall WP share in panel~(a) is much lower. Panel~(c) makes the same contrast quantitative: relative WP density at capacity is typically $2$--$4\times$, and sometimes much higher on low-baseline benches such as \texttt{lbm}, whereas LQ and SQ are mixed and are occasionally under-represented at the tip (green cells below~1). WP is therefore not merely present in these structures; it is disproportionately present in the occupancy regime that blocks the pipeline.

Figure~\ref{fig:wp_residue_cost}(a) turns that tip pressure into frontend cost. About 5\% of the region on \texttt{xalancbmk} and \texttt{x264}, and about 1.9\% on \texttt{perlbench}, is rename-blocked by a WP-filled ROB; IQ full stalls are smaller where they appear. Panel~(b) shows that this cost is structured by recovery episodes: on \texttt{x264} a single squash epoch carries about half of the WP-ROB stall and the top ten about 86\%, whereas other high-stall benches spread the same kind of stall across many epochs. Panels~(c) and~(d) are the memory complement rather than another stall metric. Among benches with enough WP demand opens, most WP-opened demand MSHRs later receive a CP coalesce (typically about 65--90\%), and on several benches about 10--24\% of CP demand joins land on a WP-opened entry (about 24\% on \texttt{xalancbmk}). WP residue in the ROB or IQ is therefore a CP cost at rename, while WP residue in a demand MSHR can still provide useful in-flight fill for CP after the opener is squashed. Occupancy share, tip density, stall cost, epoch concentration, and MSHR coalesce emphasize overlapping but not identical benches, so the footprint and its consequences are complementary; all of them require the join from occupancy holders to squash epochs.

\paragraph{Claim.} 
MFIR's contribution is that join: it identifies which entity held a shared slot, whether that entity was discarded with a WP recovery, which CP cost followed at the frontend, which recovery episodes dominate that cost, and whether a CP demand later coalesced onto a WP-opened MSHR. Figures~\ref{fig:wp_residue_occupancy} and~\ref{fig:wp_residue_cost} report the resulting occupancy footprint and its cost and reuse consequences. Those are the quantities an architect needs before choosing a structure size or reclaim policy, or before interpreting WP demand traffic, and they are the quantities that aggregate WP-blind analyses cannot produce.

\subsection{Generality Across Simulators}
\label{sec:eval:xsim}

As demonstrated in our case studies, Microflow natively supports distinct simulators like gem5~\cite{lowe2020gem5} and ChampSim~\cite{gober2022championship}. To strictly quantify cross-simulator generality, we extend our evaluation to include Scarab~\cite{oh2024udp}. 

Because trace-based and execution-driven models ingest instructions differently, we route a shared binary through three distinct frontends: Intel Pin~\cite{luk2005pin} for ChampSim, DynamoRIO~\cite{dynamorio2018tool} for Scarab, and direct syscall emulation for gem5. This guarantees identical retired instruction streams across 2M instructions and 13 distinct program counters, verified by matching hashes directly from the compiled MFIR. Applying a single unmodified analysis script across all three platforms yields nearly identical results: measured cycles exhibit a negligible 0.0002\% spread around the 1.6M cycle baseline, uniformly deriving a 1.25 IPC and accurately recovering the 192-entry ROB capacity from tenancy intervals. 

When the generated representations do diverge, they strictly reflect expected taxonomic variations between codebases. For example, ChampSim logs the primary stall between issue and execute, whereas Scarab and gem5 record the exact same stall between decode and issue due to differing internal definitions of the issue stage. Our representation natively resolves these variations using per-architecture profiles that map simulator-specific event vocabularies into a unified structural abstraction. This establishes a strict separation of concerns: porting to a new simulator simply requires defining a localized profile that acts as a contract, translating the target simulator's unique vocabulary into the shared event space. Ultimately, this decoupling empowers architects to author complex queries once and deploy them seamlessly, entirely eliminating the burden of maintaining bespoke scripts for every new environment.

\subsection{Overhead and Scalability}
\label{sec:eval:overhead}

\begin{figure}[ht]
  \centering
  \includegraphics[width=\linewidth]{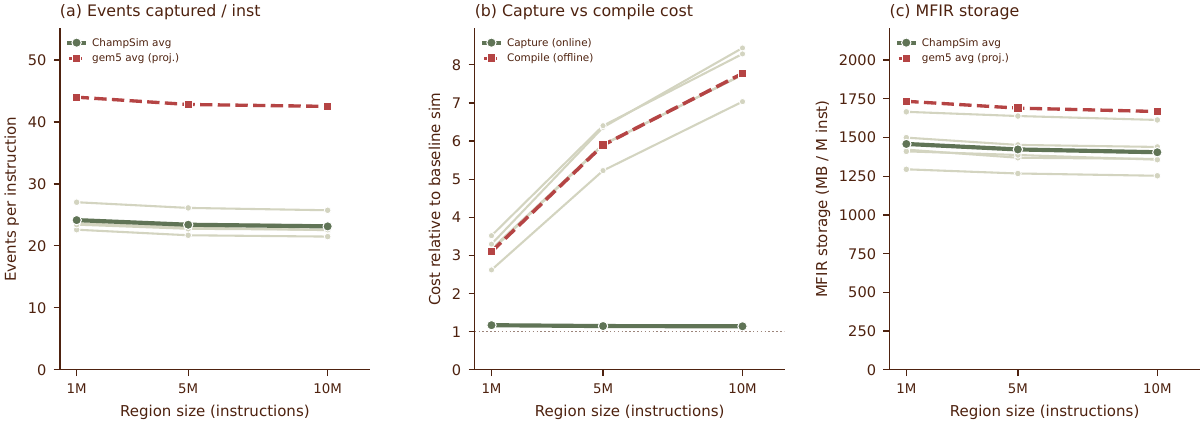}
  \caption{Capture and MFIR compile cost vs. ROI length. (a) Near-flat event density. (b) Online capture 1.16x baseline; offline compile separate. (c) Near-constant MFIR storage density (3x trace file)}
  \label{fig:overhead}
\end{figure}

Figure~\ref{fig:overhead} separates Microflow's cost into what runs \emph{inside} the simulator and what runs \emph{after} it, across ROI lengths of 1M, 5M, and 10M instructions. Panel~(a) shows that event density is near-flat (ChampSim suite mean $\approx$24events/instruction; gem5 denser where measured), so evidence volume grows essentially linearly with the window. Panel(b) contrasts the two wall-clock terms relative to an uninstrumented baseline: online capture stays near $1.16\times$ at every size. Panel~(c) shows MFIR storage density likewise near-constant ($\approx$3$\times$ the trace on ChampSim), so total artifact size scales linearly.
\section{Related Works}
\label{sec:related}
Frameworks like Top-Down Analysis (TMA) classify pipeline slots using hardware performance counters~\cite{yasin2014top}, while per-instruction profilers (PICS) such as TEA~\cite{gottschall2023tea} and DIP~\cite{desantana2026dip} establish cycle stacks anchored at commit or dispatch. While these methods provide useful breakdowns, their attribution vocabulary is closed and instruction-centric. Treating the instruction as the sole unit of analysis leaves non-instruction entities like cache packets, prefetch requests, MSHR entries, and wrong-path occupancies unnameable, flattening complex hardware interactions into instruction-local symptoms.

Approaches like icost~\cite{fields2003interaction}, Calipers~\cite{golestani2022calipers}, and ArchExplorer~\cite{bai2023archexplorer} model microexecution as directed acyclic graphs to evaluate critical paths and resource bottlenecks. However, these graphs rely on heuristic rules to infer edges from traces, leading to significant quantified errors in critical-path and resource estimations~\cite{bai2023archexplorer}. Furthermore, because their vertices are indexed by instruction stages, non-instruction objects lack native representation.

Unlike aggregate counters, sampled profiling, or heuristic graphs that attempt to infer dependencies offline, \Name{} records causal evidence precisely where actions are decided during simulation. By compiling the event stream into the \Name{} intermediate representation (MFIR), \Name{} replaces inferred edges with observed, capture-time causal relations. By decoupling simulation capture from relational analysis, MFIR's schema naturally accommodates alternative formulations such as PICS-style projections, and dependence-graph traversals as queries over a single artifact.
\section{Discussion and Future Work}
\label{sec:discuss}

Microflow provides a unified foundation for cross-layer microarchitectural analysis, but its full potential extends beyond the capabilities demonstrated in this work. The framework opens several promising avenues for future research and system evolution.

\textbf{\textit{AI-Assisted Analysis and Agentic Design Exploration.}} 
Recent advances in tool-using AI agents highlight the feasibility of autonomous agents that coordinate heterogeneous tools, execute multi-step analytical workflows, and synthesize high-level reasoning. These capabilities are beginning to influence systems and architecture research~\cite{cheng2025barbarians, yazdanbakhsh2025beyond, gupta2026archagent}, where AI-driven agents are increasingly explored for performance diagnosis, design-space exploration, and automated insight generation. Microflow is well-positioned to benefit from similar integration. At the simplest level, Natural-Language-to-Query (NL2Query) agents~\cite{liu2025survey, zhang2025structureguided} can translate user prompts into executable analytical queries issued directly to the analysis engine, dramatically lowering the expertise required to extract insights. More advanced agents could orchestrate multi-step workflows that combine exploratory queries and analysis backend execution. Because Microflow exposes a consistent, typed query interface and a well-defined analysis API, these agents can operate across diverse workloads and datasets using defined tools over this analysis backend.

\textbf{\textit{Extending Microflow Beyond CPUs.}}
While the current implementation targets CPU simulators, the MFIR abstractions are designed to extend beyond CPUs. GPU simulators~\cite{khairy2020accel, sun19mgpusim}, memory-system simulators~\cite{luo2023ramulator, li2020dramsim3}, and ML accelerators~\cite{raj2025scale} can all serve as simulator frontends.


\section{Concluding Remarks}
\label{sec:conclude}

Instruction-centric models impose a fundamental ceiling on microarchitectural analysis by reducing complex cross-entity interactions to isolated downstream symptoms. To eliminate this blind spot, we introduced Microflow, a framework realizing causal observability. By combining microarchitectural abstractions into an intermediate representation (MFIR), Microflow decouples simulation capture from causal analysis. This transforms complex diagnostics into fast queries without writing bespoke analysis scripts and costly re-simulations for each new question. Our evaluations across diverse benchmarks and simulators show that Microflow successfully resolves hidden pathologies, from decomposing prefetcher oracle headroom to quantifying wrong-path pipeline blockages.



\bibliographystyle{ACM-Reference-Format}
\bibliography{references}

@book{hennessy-90,
  author    = {John L. Hennessy and David A. Patterson},
  title     = {Computer Architecture: A Quantitative Approach},
  year      = {1990},
  publisher = {Morgan Kaufmann Publishers Inc.},
  isbn      = {1-55880-069-8}
}

@inproceedings{fields2003interaction,
  title     = {Using Interaction Costs for Microarchitectural Bottleneck Analysis},
  author    = {Fields, Brian A. and Bod{\'\i}k, Rastislav and Hill, Mark D. and Newburn, Chris J.},
  booktitle = {36th Annual IEEE/ACM International Symposium on Microarchitecture (MICRO-36)},
  pages     = {228--239},
  year      = {2003},
  publisher = {IEEE}
}

@inproceedings{lee2014rpstacks,
  title={Rpstacks: Fast and accurate processor design space exploration using representative stall-event stacks},
  author={Lee, Jaewon and Jang, Hanhwi and Kim, Jangwoo},
  booktitle={2014 47th Annual IEEE/ACM International Symposium on Microarchitecture},
  pages={255--267},
  year={2014},
  organization={IEEE}
}

@article{tanimoto2017dependence,
  title={Dependence graph model for accurate critical path analysis on out-of-order processors},
  author={Tanimoto, Teruo and Ono, Takatsugu and Inoue, Koji},
  journal={Journal of Information Processing},
  volume={25},
  pages={983--992},
  year={2017},
  publisher={Information Processing Society of Japan}
}

@article{golestani2019performance,
  title={Performance Modeling and Bottleneck Analysis of EDGE Processors Using Dependence Graphs},
  author={Golestani, Hossein and Gupta, Gagan and Sen, Rathijit},
  journal={IEEE Computer Architecture Letters},
  volume={18},
  number={1},
  pages={79--82},
  year={2019},
  publisher={IEEE}
}

@inproceedings{golestani2022calipers,
  title     = {Calipers: A Criticality-aware Framework for Modeling Processor Performance},
  author    = {Golestani, Hossein and others},
  booktitle = {36th ACM International Conference on Supercomputing (ICS)},
  year      = {2022}
}

@inproceedings{bai2023archexplorer,
  title     = {ArchExplorer: Microarchitecture Exploration via Bottleneck Analysis},
  author    = {Bai, Chen and others},
  booktitle = {56th Annual IEEE/ACM International Symposium on Microarchitecture (MICRO)},
  year      = {2023}
}

@inproceedings{zheng2023ccs,
  title={CCS: A Motif-Based Storage Format for Micro-execution Dependence Graph},
  author={Zheng, Yawen and Han, Chenji and Zhang, Tingting and Yang, Chao and Wang, Jian},
  booktitle={International Symposium on Advanced Parallel Processing Technologies},
  pages={130--146},
  year={2023},
  organization={Springer}
}

@article{zheng2024dependence,
  title={A dependence graph pattern mining method for processor performance analysis},
  author={Zheng, Yawen and Han, Chenji and Zhang, Tingting and Zhang, Fuxin and Wang, Jian},
  journal={Performance Evaluation},
  volume={164},
  pages={102409},
  year={2024},
  publisher={Elsevier}
}

@inproceedings{yasin2014top,
  title={A top-down method for performance analysis and counters architecture},
  author={Yasin, Ahmad},
  booktitle={2014 IEEE International Symposium on Performance Analysis of Systems and Software (ISPASS)},
  pages={35--44},
  year={2014},
  organization={IEEE}
}

@article{eyerman2006performance,
  title={A performance counter architecture for computing accurate CPI components},
  author={Eyerman, Stijn and Eeckhout, Lieven and Karkhanis, Tejas and Smith, James E},
  journal={ACM SIGPLAN Notices},
  volume={41},
  number={11},
  pages={175--184},
  year={2006},
  publisher={ACM New York, NY, USA}
}

@inproceedings{de2026fyi,
  title={FYI: A Foundational Counter Architecture for Building CPI Stacks on Out-of-Order Processors},
  author={de Santana, Silvio Campelo and Eeckhout, Lieven and Jahre, Magnus},
  booktitle={2026 IEEE International Symposium on Performance Analysis of Systems and Software (ISPASS)},
  pages={358--368},
  year={2026},
  organization={IEEE}
}

@article{gottschall2024per,
  title={Per-instruction cycle stacks through time-proportional event analysis},
  author={Gottschall, Bj{\"o}rn and Eeckhout, Lieven and Jahre, Magnus},
  journal={IEEE Micro},
  volume={44},
  number={4},
  pages={27--33},
  year={2024},
  publisher={IEEE}
}

@inproceedings{gottschall2023tea,
  title     = {TEA: Time-Proportional Event Analysis},
  author    = {Gottschall, Bj{\"o}rn and Eeckhout, Lieven and Jahre, Magnus},
  booktitle = {50th Annual International Symposium on Computer Architecture (ISCA)},
  year      = {2023}
}

@inproceedings{desantana2026dip,
  title     = {Chips Need DIP: Time-Proportional Per-Instruction Cycle Stacks at Dispatch},
  author    = {Campelo de Santana, Silvio Heverton and Rogers, Joseph and Eeckhout, Lieven and Jahre, Magnus},
  booktitle = {31st ACM International Conference on Architectural Support for Programming Languages and Operating Systems (ASPLOS), Volume 2},
  pages     = {361--376},
  year      = {2026}
}

@inproceedings{de2026pesto,
  title={Pesto: Diagnosing Performance Pathologies in Out-of-Order Processors},
  author={de Santana, Silvio Campelo and Rogers, Joseph and Eeckhout, Lieven and Jahre, Magnus},
  booktitle={2026 IEEE International Symposium on Performance Analysis of Systems and Software (ISPASS)},
  pages={405--415},
  year={2026},
  organization={IEEE}
}

@article{lowe2020gem5,
  title={The gem5 simulator: Version 20.0+},
  author={Lowe-Power, Jason and Ahmad, Abdul Mutaal and Akram, Ayaz and Alian, Mohammad and Amslinger, Rico and Andreozzi, Matteo and Armejach, Adri{\`a} and Asmussen, Nils and Beckmann, Brad and Bharadwaj, Srikant and others},
  journal={arXiv preprint arXiv:2007.03152},
  year={2020}
}

@article{gober2022championship,
  title={The championship simulator: Architectural simulation for education and competition},
  author={Gober, Nathan and Chacon, Gino and Wang, Lei and Gratz, Paul V and Jimenez, Daniel A and Teran, Elvira and Pugsley, Seth and Kim, Jinchun},
  journal={arXiv preprint arXiv:2210.14324},
  year={2022}
}

@inproceedings{khairy2020accel,
  title={Accel-sim: An extensible simulation framework for validated gpu modeling},
  author={Khairy, Mahmoud and Shen, Zhesheng and Aamodt, Tor M and Rogers, Timothy G},
  booktitle={2020 ACM/IEEE 47th Annual International Symposium on Computer Architecture (ISCA)},
  pages={473--486},
  year={2020},
  organization={IEEE}
}

@article{luo2023ramulator,
  title={Ramulator 2.0: A modern, modular, and extensible dram simulator},
  author={Luo, Haocong and Tu{\u{g}}rul, Yahya Can and Bostanc{\i}, F Nisa and Olgun, Ataberk and Ya{\u{g}}l{\i}k{\c{c}}{\i}, A Giray and Mutlu, Onur},
  journal={IEEE Computer Architecture Letters},
  volume={23},
  number={1},
  pages={112--116},
  year={2023},
  publisher={IEEE}
}

@article{li2020dramsim3,
  title={DRAMsim3: A cycle-accurate, thermal-capable DRAM simulator},
  author={Li, Shang and Yang, Zhiyuan and Reddy, Dhiraj and Srivastava, Ankur and Jacob, Bruce},
  journal={IEEE Computer Architecture Letters},
  volume={19},
  number={2},
  pages={106--109},
  year={2020},
  publisher={IEEE}
}

@inproceedings{raj2025scale,
  title={SCALE-Sim v3: A modular cycle-accurate systolic accelerator simulator for end-to-end system analysis},
  author={Raj, Ritik and Banerjee, Sarbartha and Chandra, Nikhil and Wan, Zishen and Tong, Jianming and Samajdhar, Ananda and Krishna, Tushar},
  booktitle={2025 IEEE International Symposium on Performance Analysis of Systems and Software (ISPASS)},
  pages={186--200},
  year={2025},
  organization={IEEE}
}

@misc{perais2018cvp1,
  author       = {Perais, A.},
  title        = {{Championship Value Prediction 1 Public Traces}},
  year         = {2018},
  organization = {Univ. Grenoble Alpes, CNRS, Grenoble INP, TIMA},
  howpublished = {\url{https://perscido.univ-grenoble-alpes.fr/datasets/DS382}}
}

@misc{intelVTune,
  author       = {{Intel Corporation}},
  title        = {Intel® VTune™ Profiler},
  year         = {2025},
  howpublished = {\url{https://www.intel.com/content/www/us/en/developer/tools/oneapi/vtune-profiler.html}},
  note         = {Accessed October 2025}
}

@misc{mundichipparakkal2025arm,
  title={Arm Neoverse V1 Core: Performance Analysis Methodology. White Paper. Arm Limited},
  author={Mundichipparakkal, Jumana and Cai, Yibo and Ellis, Sam and Greene, Darin and Williams, Michael and Chouhan, Tushar},
  institution = {Arm Limited},
  year={2025},
  url         = {https://developer.arm.com/documentation/109199/latest/}
}

@manual{amd2024uprof,
  organization = {AMD Advanced Micro Devices},
  title        = {AMD uProf User Guide},
  edition      = {Revision 4.2},
  year         = {2024},
  month        = {January},
  note         = {Publication Number: 57368}
}

@inproceedings{wang2026r,
  title={R-Max: Extending B{\'e}L{\'a}Dy's MIN with Prefetching to Bound Realistic Cache Performance},
  author={Wang, Lei and Lee, Chia-Hang and Merrell, Maccoy and Chacon, Gino and Jim{\'e}nez, Daniel A and Gratz, Paul V},
  booktitle={2026 ACM/IEEE 53rd Annual International Symposium on Computer Architecture (ISCA)},
  pages={2270--2286},
  year={2026},
  organization={IEEE}
}

@inproceedings{kim2016path,
  title={Path confidence based lookahead prefetching},
  author={Kim, Jinchun and Pugsley, Seth H and Gratz, Paul V and Reddy, AL Narasimha and Wilkerson, Chris and Chishti, Zeshan},
  booktitle={2016 49th Annual IEEE/ACM International Symposium on Microarchitecture (MICRO)},
  pages={1--12},
  year={2016},
  organization={IEEE}
}

@article{de2012open,
  title={Open Trace Format 2 The Next Generation of Scalable Trace Formats and Support Libraries},
  author={De Bosschere, K and others},
  journal={Applications, Tools and Techniques on the Road to Exascale Computing},
  volume={22},
  pages={481},
  year={2012},
  publisher={IOS Press}
}

@misc{opentelemetry,
  author       = {{OpenTelemetry Authors}},
  title        = {OpenTelemetry},
  year         = {2026},
  howpublished = {\url{https://opentelemetry.io/}},
  note         = {Accessed: 2026-09-08}
}

@article{systemctlm,
  author={},
  journal={IEEE Std 1666-2023 (Revision of IEEE Std 1666-2011)}, 
  title={IEEE Standard for Standard SystemC® Language Reference Manual}, 
  year={2023},
  volume={},
  number={},
  pages={1-618},
  doi={10.1109/IEEESTD.2023.10246125}}

@inproceedings{lattner2004llvm,
  title={LLVM: A compilation framework for lifelong program analysis \& transformation},
  author={Lattner, Chris and Adve, Vikram},
  booktitle={International symposium on code generation and optimization, 2004. CGO 2004.},
  pages={75--86},
  year={2004},
  organization={IEEE}
}

@inproceedings{lattner2021mlir,
  title={MLIR: Scaling compiler infrastructure for domain specific computation},
  author={Lattner, Chris and Amini, Mehdi and Bondhugula, Uday and Cohen, Albert and Davis, Andy and Pienaar, Jacques and Riddle, River and Shpeisman, Tatiana and Vasilache, Nicolas and Zinenko, Oleksandr},
  booktitle={2021 IEEE/ACM international symposium on code generation and optimization (CGO)},
  pages={2--14},
  year={2021},
  organization={IEEE}
}

@inproceedings{wunderlich2003smarts,
  title={SMARTS: Accelerating microarchitecture simulation via rigorous statistical sampling},
  author={Wunderlich, Roland E and Wenisch, Thomas F and Falsafi, Babak and Hoe, James C},
  booktitle={Proceedings of the 30th annual international symposium on Computer architecture},
  pages={84--97},
  year={2003}
}

@article{hamerly2005simpoint,
  title={Simpoint 3.0: Faster and more flexible program phase analysis},
  author={Hamerly, Greg and Perelman, Erez and Lau, Jeremy and Calder, Brad},
  journal={Journal of Instruction Level Parallelism},
  volume={7},
  number={4},
  pages={1--28},
  year={2005}
}

@misc{perfetto,
  author       = {{Google}},
  title        = {Perfetto},
  howpublished = {\url{https://perfetto.dev/}},
  note         = {Accessed: September 2026},
  year         = {2026}
}

@article{luk2005pin,
  title={Pin: building customized program analysis tools with dynamic instrumentation},
  author={Luk, Chi-Keung and Cohn, Robert and Muth, Robert and Patil, Harish and Klauser, Artur and Lowney, Geoff and Wallace, Steven and Reddi, Vijay Janapa and Hazelwood, Kim},
  journal={Acm sigplan notices},
  volume={40},
  number={6},
  pages={190--200},
  year={2005},
  publisher={ACM New York, NY, USA}
}

@misc{dynamorio2018tool,
  author       = {Q. Zhao and Derek L. Bruening},
  title        = {DynamoRIO: Dynamic Instrumentation Tool Platform},
  howpublished = {\url{https://github.com/DynamoRIO/dynamorio}},
  note         = {Accessed: October 2025},
  year         = {2025}
}

@inproceedings{oh2024udp,
  author = {Oh, Surim and Xu, Mingsheng and Khan, Tanvir Ahmed and Kasikci, Baris and Litz, Heiner},
  title = {UDP: Utility-Driven Fetch Directed Instruction Prefetching},
  booktitle = {Proceedings of the 51st International Symposium on Computer Architecture (ISCA)},
  series = {ISCA 2024},
  year = {2024},
  month = jun,
}

@inproceedings{sun19mgpusim, 
    author = {Sun, Yifan and Baruah, Trinayan and Mojumder, Saiful A. and Dong, Shi and Gong, Xiang and Treadway, Shane and Bao, Yuhui and Hance, Spencer and McCardwell, Carter and Zhao, Vincent and Barclay, Harrison and Ziabari, Amir Kavyan and Chen, Zhongliang and Ubal, Rafael and Abell\'{a}n, Jos\'{e} L. and Kim, John and Joshi, Ajay and Kaeli, David}, 
    title = {MGPUSim: Enabling Multi-GPU Performance Modeling and Optimization}, 
    year = {2019}, 
    isbn = {9781450366694}, 
    publisher = {Association for Computing Machinery}, 
    address = {New York, NY, USA}, 
    url = {https://doi.org/10.1145/3307650.3322230}, 
    doi = {10.1145/3307650.3322230}, 
    booktitle = {Proceedings of the 46th International Symposium on Computer Architecture}, 
    pages = {197–209}, 
    numpages = {13}, 
    location = {Phoenix, Arizona}, 
    series = {ISCA '19} 
}

@article{cheng2025barbarians,
  title={Barbarians at the Gate: How AI is Upending Systems Research},
  author={Cheng, Audrey and Liu, Shu and Pan, Melissa and Li, Zhifei and Wang, Bowen and Krentsel, Alex and Xia, Tian and Cemri, Mert and Park, Jongseok and Yang, Shuo and others},
  journal={arXiv preprint arXiv:2510.06189},
  year={2025}
}

@article{gupta2026archagent,
  title={ArchAgent: Agentic AI-driven Computer Architecture Discovery},
  author={Gupta, Raghav and Jain, Akanksha and Gonzalez, Abraham and Novikov, Alexander and Huang, Po-Sen and Balog, Matej and Eisenberger, Marvin and Shirobokov, Sergey and V{\~u}, Ng{\^a}n and Dixon, Martin and others},
  journal={arXiv preprint arXiv:2602.22425},
  year={2026}
}

@article{yazdanbakhsh2025beyond,
  title={Beyond Moore's Law: Harnessing the Redshift of Generative AI with Effective Hardware-Software Co-Design},
  author={Yazdanbakhsh, Amir},
  journal={arXiv preprint arXiv:2504.06531},
  year={2025}
}

@article{liu2025survey,
  title={A Survey of Text-to-SQL in the Era of LLMs: Where are we, and where are we going?},
  author={Liu, Xinyu and Shen, Shuyu and Li, Boyan and Ma, Peixian and Jiang, Runzhi and Zhang, Yuxin and Fan, Ju and Li, Guoliang and Tang, Nan and Luo, Yuyu},
  journal={IEEE Transactions on Knowledge and Data Engineering},
  year={2025},
  publisher={IEEE}
}

@inproceedings{zhang2025structureguided,
title={Structure-Guided Large Language Models for Text-to-{SQL} Generation},
author={Qinggang Zhang and Hao Chen and Junnan Dong and Shengyuan Chen and Feiran Huang and Xiao Huang},
booktitle={Forty-second International Conference on Machine Learning},
year={2025},
url={https://openreview.net/forum?id=gT8JSEFqaS}
}

@misc{SPEC2017,
    title = {SPEC. Standard Performance Evaluation Corporation
SPEC CPU 2017.},
    url = {spec:org/cpu2017/},
    urldate = {2024-06-05},
    year = {2017}
}

\end{document}